\documentclass[11pt,aps,pra,tightenlines,nofootinbib,longbibliography]{revtex4-2}
\pdfoutput=1

\usepackage{amsmath}
\usepackage{amssymb}
\usepackage{bbm}
\usepackage{simplewick}
\usepackage{mathtools}
\usepackage[normalem]{ulem}
\usepackage{dsfont}
\usepackage{mathrsfs}
\usepackage[makeroom]{cancel}
\usepackage{graphicx}
\usepackage{bm}

\def\rhat{{\hat {\boldsymbol r}}}
\def\khat{{\hat {\boldsymbol k}}}

\renewcommand{\sinh}{\operatorname{sh}}
\renewcommand{\cosh}{\operatorname{ch}}

\def\0{\boldsymbol{0}}
\def\RR{\mathbbm{R}}

\def\xx{\boldsymbol{x}}
\def\kk{{\boldsymbol{k}}}
\def\rr{\boldsymbol{r}}
\def\yy{{\boldsymbol{y}}}

\newcommand{\Rzymskie}[1]{%
  \textup{\uppercase\expandafter{\romannumeral#1}}%
}

\def\rmJ{{\rm J}}

\def\rmH{{\rm H}}

\def\J{\boldsymbol{J}}
\def\V{\boldsymbol{V}}
\def\E{\boldsymbol{E}}

\def\Bold{\boldsymbol{B}}

\def\Lag{{\cal L}}
\def\ii{\mathrm{i}}

\def\dd#1{d^3\mkern-1.5mu#1\,}

\def\bnabla{{\boldsymbol\nabla}}

\def\boldeta{{\boldsymbol\eta}}
\def\boldmu{{\boldsymbol\mu}}

\def\vareps#1{\varepsilon_{#1}}
\def\om#1{\omega_{#1}}

\def\smallint{{\textstyle\int}}

\def\B#1{\left(#1\right)}
\def\BB#1{\left[#1\right]}

\def\la{\langle}
\def\ra{\rangle}

\def\lara#1{\la#1\ra}

\def\be{\begin{equation}}
\def\ee{\end{equation}}

\def\hc{\text{h.c.}}
\def\cc{\text{c.c.}}

\def\for{\ \text{for} \ }

\usepackage[ddmmyyyy]{datetime}
\makeatletter
\def\Dated@name{}
\makeatother

\def\DIV{\text{div}}
\def\ddd{\boldsymbol{d}}

\def\pref{0.575}

\begin{document}
\title{Non-equilibrium dynamics of 
 dipole-charged fields in the Proca theory}
\author{Bogdan Damski}
\affiliation{Jagiellonian University, 
Faculty of Physics, Astronomy and Applied Computer Science,
{\L}ojasiewicza 11, 30-348 Krak\'ow, Poland}
\begin{abstract}
We discuss the dynamics of 
field configurations encoded in 
the certain class
of electric (magnetic) dipole-charged states in 
the Proca theory of the real massive vector field.
We construct such states   so as to ensure 
that the long distance structure  of the mean 
electromagnetic  field in them 
is initially set by the formula  describing the 
electromagnetic field of the electric (magnetic)
dipole. We analyze  then how such a
mean electromagnetic field 
evolves in time. 
We find that far away from the center of the initial field
configuration, the long range 
component of the mean electromagnetic field 
harmonically oscillates, which  
leads to the phenomenon of 
the periodic   oscillations
of the electric (magnetic) dipole moment.
We also find that near the center of the initial field
configuration,
the mean 
electromagnetic field 
 escapes  from its  initial 
 arrangement and a spherical 
 shock wave propagating with the 
 speed of light appears in the studied system. 
 A curious configuration   
 of the axisymmetric  mean electric field
is found to  accompany   the
 mean magnetic field in magnetic 
dipole-charged states.
\end{abstract}
\maketitle

\section{Introduction}
\label{Introduction_sec}

The  Proca theory studied in this work
is defined by the 
following 
Lagrangian density  
\cite{Greiner,ColemanBook,Weinberg}
\be
\Lag=-\frac{1}{4} 
(\partial_\mu V_\nu-\partial_\nu V_\mu)^2
+\frac{m^2}{2}(V_\mu)^2,
\label{SPr}
\ee
where 
 $V$ is the vector field operator 
 and $m>0$ is the mass of the vector 
boson that it describes 
(see Appendix \ref{Conv_app}  for our conventions
and \cite{Nieto_RMP1971,Gillies2005,Nieto_RMP2010} for the general 
discussion of the Proca theory in the context of massive 
photon  electrodynamics).
Such a theory differs from the 
Maxwell theory by the mass term, which 
leads to the non-vanishing energy of 
   small momentum
excitations  
\be
\vareps{k}=\sqrt{m^2+|\kk|^2}= m + O\B{|\kk|^2} \for 
\kk\to\0.
\label{ek00}
\ee
This simple observation implies 
that the  mass $m$
 fundamentally impacts  the
large-distance properties  of
the  electromagnetic field
of the Proca theory,
which is
represented
by the operators
$\E=-\partial_0\V-\bnabla V^0$
and $\Bold=\bnabla \times \V$.

As far as the equilibrium 
configurations of such fields are 
of interest, the consequences of 
$m\neq0$ are well known.
To illustrate  some of them, 
we consider for a moment  the classical Proca  
field coupled to an external $4$-current.
The formulas describing the electromagnetic 
field of the 
point charge and the  point magnetic 
dipole in the Proca theory 
are the following.\footnote{Denoting by $J_\text{ext}$ the external 
$4$-current, we define the   point charge by 
$J_\text{ext}=(q\delta(\rr),\0)$, where
$q$ is a non-zero real parameter.
For the point  magnetic dipole, we assume that 
$J_\text{ext}=(0,\J_\text{ext})$, where 
$\J_\text{ext}$ has infinitesimally 
small spatial support and the value such that 
$\boldmu=\int d^3r\, \rr\times\J_\text{ext}/2\neq\0$.}
In the former case, 
the magnetic field vanishes
while  the  electric field is 
 given by 
\be
\E_\text{cl}=-\bnabla\B{\frac{q\exp(-mr)}{4\pi r}}
\for r=|\rr|>0,
\label{Eclcl1}
\ee
where the subscript $\text{cl}$ reminds 
us that we discuss classical expressions now (see e.g. 
\cite{Gillies2005,Nieto_RMP2010,Hertzberg2019}).
In the latter case, the electric field vanishes
while the magnetic  field is given by
(see e.g. \cite{Nieto_RMP1971,Gillies2005})
\be
\Bold_\text{cl}=\bnabla\times\BB{\bnabla\times\B{\frac{\boldmu\exp(-mr)}{4\pi r}}}
\for r>0.
\label{Bclcl1}
\ee
As is explained in \cite{Nieto_RMP1971,Gillies2005,Nieto_RMP2010},
(\ref{Eclcl1}), (\ref{Bclcl1}), and related expressions were critically 
compared to  experimental data (e.g. obtained during 
the studies of  the Coulomb's law
and the  magnetic field of Earth). 
Such comparisons lead to 
the determination of various upper 
bounds on the photon mass.
In the context of our work,
(\ref{Eclcl1}) and (\ref{Bclcl1})
illustrate the fact that 
the mass $m$ introduces exponential damping of 
{\it equilibrium}
electromagnetic field configurations, 
which makes such 
 field configurations short-range
 in character.

We are interested  in the impact 
of the mass $m$ on 
long-range electromagnetic 
 fields in the Proca theory.
 As the above discussion suggests,
 such fields do not represent
 equilibrium field configurations.
  However, switching to the quantum theory from now on, 
  we would like to stress that finite-energy normalizable states, 
  representing long-range electromagnetic fields,   can be 
  constructed in  the  Proca  theory.
The studies of their non-equilibrium dynamics 
have been initiated in 
our recent
work \cite{BDPeriodic1}, which we briefly
summarize below to set the stage for the
presentation of our follow up
results.

\begin{figure}[t]
\includegraphics[width=\pref\columnwidth,clip=true]{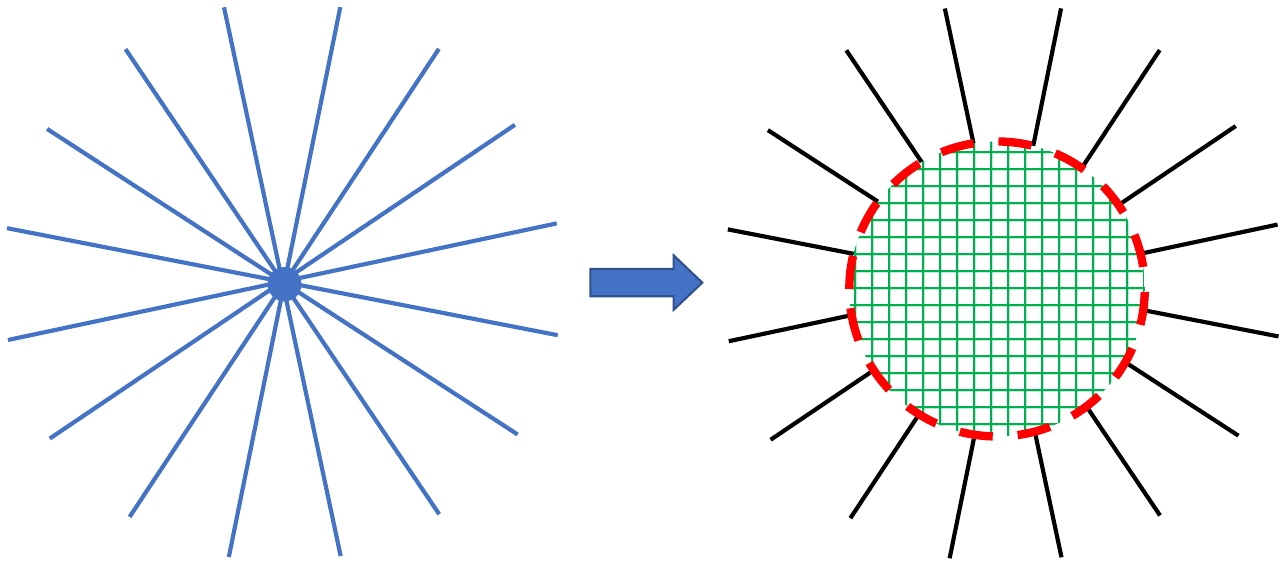}
\caption{Schematic illustration of the 
non-equilibrium 
dynamics of the mean electric field 
in the states discussed in \cite{BDPeriodic1}. 
The  plots  depict the initial 
mean  electric field, approximately
given by the  Coulomb formula, 
and its non-equilibrium configuration 
after  some time evolution.
The dashed red line shows the shock wave
front that is localized on a sphere,
whereas the crossed lines 
show the region of space that 
has already been swept  by  
the shock wave. 
 The mean electric field also 
 evolves  outside the 
 shock wave sphere,  where it 
 is dominated by the 
 periodic oscillations 
 of its Coulomb component.
 This is illustrated by the 
 different colors  of the 
 field lines in the two 
 plots.
}
 \label{fig1}
\end{figure}

Namely, we  studied the dynamics of 
the mean electric field in the certain class of
charged states in the Proca theory \cite{BDPeriodic1}.
The hallmark feature  of such states  was 
 the non-zero 
 expectation value of the charge operator
 \be
Q(t)=\int \dd{y} \DIV\E(t,\yy)
\label{QdivE}
\ee
in them.
This was analyzed  in the following 
setup (Fig. \ref{fig1}). 
At some time, say $t=0$, we assumed that 
the system was  in the state,
where  the mean electric field 
was   asymptotically  given by 
the Coulomb formula
\be
\left.\lara{\E}\right|_{t=0}\propto \frac{\rr}{4\pi r^3}
\for r\to\infty.
\label{Coulomb_ini}
\ee
For $t>0$, the  
mean electric field underwent 
non-equilibrium dynamics 
 because there was
no external source generating 
field (\ref{Coulomb_ini}) in the 
studied system. This lead to 
the appearance of the shock wave
that  was localized 
 on the expanding sphere of radius $t$
 (the mean electric field 
 was weakly discontinuous at $r=t$).\footnote{We use the term weakly discontinuous 
      to  refer to continuous 
      physical  quantities that 
       have discontinuous either  first or  higher order 
      derivative(s).}
Space in the considered problem was split into
two regions. The one that  had  already been  
swept  by the shock wave ($r<t$) and the one 
where the shock wave had not  yet arrived  ($r>t$).
In the former region,
the dynamics of the mean electric
field was  non-universal because  it depended
on the short distance properties  
of such a  field at $t=0$,
which    can  be chosen in 
different ways.\footnote{Departures
from the Coulomb formula 
around the origin were   necessary for
having a finite-energy field configuration.}
In the latter region, 
 the universal feature of 
 the mean electric field was  identified.
Namely, the periodically oscillating 
Coulomb field,
\be
\propto\frac{\rr}{4\pi r^3}\cos(mt),
\label{Coulomb_t}
\ee
 dominated  the long distance 
 behavior of the mean electric field 
in the studied states.
 This result explained 
the phenomenon of periodic
 charge oscillations in the Proca theory
 \cite{BDPeriodic1}.\footnote{The technical reason for 
              periodic charge oscillations is 
	      that  charge operator 
	      (\ref{QdivE}) satisfies the 
	      harmonic oscillator equation
	      in the Proca theory. 
	      This was noted in \cite{GHK1964,Guralnik1968}
	      but it was  not  elaborated any further in these
	      publications.
	      The detailed discussion
              of charge operator
              (\ref{QdivE}), as well as  physics 
	      associated with it, is presented in  \cite{BDPeriodic1}.
	      }
A similar phenomenon was mentioned  in a different 
physical context in \cite{Hertzberg2020}.

The problem  explored  in this work 
is  related to the one, which 
has just been  explained.
Namely, we will  discuss the dynamics 
of the mean electromagnetic field in
dipole-charged states in the
Proca theory, where the asymptotic 
form of  such a field
is  initially given by  
the formula
describing the electromagnetic field of 
either  the electric  
or magnetic dipole.
Such a problem, to the best of our knowledge,
has not been studied before.

The outline of this paper is the following.
Basic facts concerning the Proca theory as well 
as the technical details related to the construction 
of the dipole-charged states are presented 
in Sec. \ref{Basic_sec} and 
Appendix
 \ref{Polynomial_app}.
 The  magnetic and 
electric
dipole-charged  states are discussed in Secs. 
\ref{Magnetic_sec} and  \ref{Electric_sec},
respectively.
It is shown there how one may construct them,
so that they represent finite-energy 
field configurations, and then the 
 dynamics of the mean electromagnetic field 
 in such  states   is analyzed.
The summary of our work is presented in 
Sec. \ref{Summary_sec}, 
whereas our  conventions are listed 
in  Appendix \ref{Conv_app}.

\section{Basic equations}
\label{Basic_sec}
The vector field operator of the 
Proca theory can be 
written as \cite{Greiner}
\begin{subequations}
\be
V^\mu(x)= \int
\frac{\dd{k}}{(2\pi)^{3/2}}\frac{1}{\sqrt{2\vareps{k}}}
  \sum_{\sigma=1}^3
\eta^\mu(\kk,\sigma) a_{\kk\sigma}\exp(-\ii k\cdot x) + \hc,
\ee
where the commutators of creation and  annihilation  operators
are   ($\sigma,\sigma'=1,2,3$)
\be
 [a_{\kk\sigma},a^\dag_{\kk'\sigma'}]=\delta_{\sigma\sigma'}\delta(\kk-\kk'),
 \
[a_{\kk\sigma},a_{\kk'\sigma'}]=0,
\ee
the transverse polarization $4$-vectors  satisfy ($i,j=1,2$)
\begin{align}
&\eta(\kk,i)=\B{0,\boldeta(\kk,i)}, \ \boldeta(\kk,i)\in\RR^3, \\
& \boldeta(\kk,i)\cdot\kk=0, \
 \boldeta(\kk,i)\cdot\boldeta(\kk,j)=\delta_{ij},
\end{align}
the longitudinal polarization $4$-vector
 is given by 
\be
\eta(\kk,3)=\B{\frac{\om{k}}{ m},\frac{\vareps{k}}{ m}\khat}, \ \khat=\kk/\om{k},
\ee
\label{Vvecoperator}%
\end{subequations}
the  operators 
$a_{\kk\sigma}$
annihilate 
the vacuum state $|0\ra$,
$\om{k}=|\kk|$, and 
$k^0=\vareps{k}$
 is tacitly 
assumed in all our $d^3k$ integrals in which  
the  integrand 
depends on $k^0$.

The electric and magnetic field operators 
of the  Proca theory 
can be computed out of (\ref{Vvecoperator})
and they read
\begin{align}
\label{piE}%
&\E(x)=
\ii\int \frac{\dd{k}}{(2\pi)^{3/2}}
\sqrt{\frac{\vareps{k}}{2}}\sum_{\sigma=1}^3
\B{1-\frac{\om{k}^2}{\vareps{k}^2}\delta_{\sigma3}}
 \boldeta(\kk,\sigma) 
a_{\kk\sigma}\exp(-\ii k\cdot x) + \hc,
\\
&\Bold(x)= \ii\int
\frac{\dd{k}}{(2\pi)^{3/2}}\frac{1}{\sqrt{2\vareps{k}}}
\sum_{\sigma=1}^2
\kk\times\boldeta(\kk,\sigma)
a_{\kk\sigma}\exp(-\ii k\cdot x) + \hc
\label{piB}%
\end{align}

The quantum states of interest will be considered 
in the following form  
\be
 |\psi(\xx)\ra=\BB{\alpha -\frac{\ii}{\alpha} \chi(\xx)}|0\ra,
 \label{psichi}
\ee
where $\chi(\xx)=\chi^\dag(\xx)$ is linear in creation and annihilation 
operators while $\alpha\in\RR\setminus\{0\}$ is  constrained 
by the requirement of 
 $\lara{\psi(\xx)|\psi(\xx)}=1$, which 
leads  to 
\be
1=\alpha^2 + \frac{1}{\alpha^2} \lara{0|\chi(\xx)\chi(\xx)|0}.
\label{normGeneral}
\ee

Besides the normalizability of the wave-function,
we will also require that 
\be
{\cal H}=\lara{\psi(\xx)|H|\psi(\xx)}=
\frac{1}{\alpha^2}\lara{0|\chi(\xx)[H,\chi(\xx)]|0}<\infty,
\label{Hmean}
\ee
where $H$ is the Hamiltonian of the Proca theory
\cite{Greiner}
\be
H=\int \dd{k}\vareps{k}\sum_{\sigma=1}^3
a_{\kk\sigma}^\dag a_{\kk\sigma}.
\label{HProca}
\ee

The mean values of operators
in state (\ref{psichi}) 
will be computed via 
\be
\lara{O(t,\rr)}=\lara{\psi(\xx)|O(t,\yy)|\psi(\xx)}
=-\ii  [O(t,\yy), \chi(\xx)],
\label{Oexp}
\ee
where  $\rr=\yy-\xx$.  The time $t>0$ 
 will be assumed in inequalities involving
$t$ and the  frequently appearing equation 
$r=t$. The operator  $\chi(\xx)$ will be 
chosen such that
$\alpha$ and $\cal H$ will
be independent of $\xx$ and 
 (\ref{Oexp}) will be 
the function of $\rr$,
which the above notation suggests.

We introduce 
\be
\phi_\gamma(t,r)=\frac{1}{2\pi^2}\int_0^\infty
d\om{k} \B{\frac{ m}{\vareps{k}}}^\gamma
{\rm j}_0(\om{k}r)\cos(\vareps{k}t),
\label{phiphi}
\ee
where $\gamma$ is a positive  even number and 
${\rm j}_n$ is the spherical
Bessel function of
the first kind of 
order $n$.\footnote{The reason for choosing 
$\gamma=2,4,6,\cdots$ is purely technical.
Namely, for such values of $\gamma$, $\phi_\gamma(t,r)$
for $r\ge t$ can be
analytically evaluated via the contour integration 
technique discussed in \cite{BDPeriodic1}.
}
Function (\ref{phiphi}), 
originally considered in \cite{BDPeriodic1},
 will appear  in the subsequent sections. 
For the present work, we need to know
the following facts.

First, for $r\ge t$
\be
\phi_\gamma(t,r)=\frac{\cos(m t)-P_\gamma(mr,mt)\exp(- m r)}{4\pi r},
\label{hatphiLexp}
\ee
where $P_2(a,b)=1$ while 
$P_{4,6,8,\cdots}(a,b)$ 
can be 
obtained  via 
\be
P_\gamma(a,b)=
P_{\gamma-2}(a,0)+
\frac{a}{\gamma-2}\B{1-\frac{d}{da}}P_{\gamma-2}(a,0)
 -\int_0^b dy\int_0^y dx P_{\gamma-2}(a,x). 
\label{PPrec}
\ee
Such  a recursive formula has been derived in  Appendix \ref{Polynomial_app}
and it leads  to  
\begin{align}
&P_4(a,b)=1+\frac{a}{2}-\frac{b^2}{2},\\
&P_6(a,b)=1+\frac{5a}{8}
+\frac{a^2}{8}
-\frac{b^2}{2}
-\frac{ab^2}{4}
+\frac{b^4}{24},
\end{align}
etc.

Second, 
in the  $0<r<t$ region, 
we are unaware how 
$\phi_\gamma(t,r)$ can be 
analytically evaluated.
However, it was shown in \cite{BDPeriodic1} 
that in such a
region of space,
(\ref{phiphi}) for $\gamma=2$
can be rewritten into the form
that can be more conveniently analyzed.

Third, it was proved in  \cite{BDPeriodic1} that 
$\phi_\gamma(t,r)$ and its derivatives  up to 
 order 
$\gamma-2$ are continuous for all $r,t>0$
(circumstantial  evidence presented in 
 \cite{BDPeriodic1} suggests that the same 
 is true for the derivatives of 
order $\gamma-1$).
It was also proved there 
that at least some derivatives 
of $\phi_\gamma(t,r)$ of order $\gamma$ are 
discontinuous at $r=t$, thereby  $\phi_\gamma(t,r)$ 
is non-analytic.

\section{Magnetic dipole-charged states}
\label{Magnetic_sec}

The states studied  here will be
constructed such that 
$\lara{\E(0,\rr)}=\0$ for $r\ge0$ and
\be
\lara{\Bold(0,\rr)}=
\frac{3(\boldmu\cdot\rhat)\rhat-\boldmu}{4\pi r^3}
\label{B00}
\ee
for $r\to\infty$, where
$\boldmu$ is  the
 magnetic dipole moment  and
 $\rhat=\rr/r$.

To proceed, we  combine the following ansatz 
\be
\chi(\xx)=\int
\frac{\dd{k}}{(2\pi)^{3/2}}
\sum_{\sigma=1}^2 g_{\sigma}(\kk)
a_{\kk\sigma}\exp(\ii\kk\cdot\xx)+\hc,
\label{magdip}
\ee
where $ g_{\sigma}(\kk)\in\RR$,
with (\ref{piB}) and (\ref{Oexp}) to find that
\be
\lara{\Bold(0,\rr)}= 
\int 
\frac{\dd{k}}{(2\pi)^3}\frac{1}{\sqrt{2\vareps{k}}}
\sum_{\sigma=1}^2g_{\sigma}(\kk)
\kk\times\boldeta(\kk,\sigma)
\exp(\ii\kk\cdot\rr)
+\cc
\ee
If we substitute\footnote{For
any $\kk\neq\0$, one may always choose $\hat{g}_\sigma(\kk)\in\RR$ 
      so as to satisfy  (\ref{gmagmag}).
      This is guaranteed by the fact that  $\boldeta(\kk,\sigma=1,2)\in\RR^3$ 
      form a basis in the plane perpendicular to $\kk$. 
      }
\begin{subequations}
\begin{align}
&g_{\sigma}(\kk)= 
\sqrt{\frac{\vareps{k}}{2\om{k}^2}} \hat{g}_\sigma(\kk),
\\ 
&\sum_{\sigma=1}^2 \hat{g}_{\sigma}(\kk) \boldeta(\kk,\sigma)
=\frac{\boldmu\times\kk}{\om{k}}
\label{gmagmag}
\end{align}
\label{ggNEW}
\end{subequations}
into such an 
expression,
we    arrive at 
\be
\lara{\Bold(0,\rr)}=\int
\frac{\dd{k}}{(2\pi)^3}\frac{\kk\times(\boldmu\times\kk)}{\om{k}^2}
\exp(\ii \kk\cdot\rr).
\label{B0000}
\ee
Moreover, we find in similar manner that   $\lara{\E(0,\rr)}=\0$  
when (\ref{magdip}) is combined with (\ref{ggNEW}).
Some  remarks are in order now.

Integral  (\ref{B0000}) has a  proper 
infrared (IR) structure.
This  remark can be formally 
quantified by replacing the cross 
product in (\ref{B0000}) with 
$-\bnabla\times(\boldmu\times\bnabla)$,
where $\bnabla=\left(\partial/\partial r^i\right)$,
and then taking such a
differential operator outside 
the integral.
The resulting expression exactly matches
(\ref{B00}) for all $r>0$. However, 
such an  exchange of the order of 
differentiation and integration 
is  not permissible  due to the 
poor  ultraviolet (UV) convergence properties of 
integral  (\ref{B0000}), which brings
us to our next  remark.

Integral (\ref{B0000}) is actually UV 
non-convergent. This remark 
can be quantified with the following 
interrelated identities
\begin{align}
\label{intOmega0}
&\int d\Omega(\khat) \exp(\ii \kk\cdot\rr)
= 4\pi {\rm j}_0(\om{k}r),\\
&\int d\Omega(\khat)k^i k^j\exp(\ii \kk\cdot\rr)
= 4\pi \delta^{ij}\frac{\om{k}}{r}
{\rm j}_1(\om{k}r)-
4\pi r^i r^j\B{\frac{\om{k}}{r}}^2 
{\rm j}_2(\om{k}r),
\label{intOmega}
\end{align}
where $\Omega(\khat)$ is the solid angle
in  momentum space.
The non-convergence issue can be fixed by  inserting 
a suitable function $f(\om{k})$ under the integral 
symbol in the 
expression for $\lara{\Bold(0,\rr)}$.
 This is done via
\be
\chi(\xx)\to\chi(\xx)=
\int
\frac{\dd{k}}{(2\pi)^{3/2}}\sqrt{\frac{\vareps{k}}{2\om{k}^2}}
f(\om{k})
\sum_{\sigma=1}^2 \hat{g}_{\sigma}(\kk)
a_{\kk\sigma}\exp(\ii\kk\cdot\xx)+\hc,
\label{cohkap00}
\ee
where  real-valued $f(\om{k})$,
which we conveniently assume to be bounded on 
$(0,\infty)$, is supposed to 
satisfy  two requirements. 
First, it should approach unity 
for $\om{k}\to0$ 
 to preserve the validity 
of (\ref{B00}) in the 
$r\to\infty$ limit. 
Second, it should  vanish fast enough 
for $\om{k}\to\infty$ 
to ensure  the  UV convergence of the
expression for $\lara{\Bold(0,\rr)}$.
The function 
$f(\om{k})$ vanishing faster than 
$1/\om{k}$ for  $\om{k}\to\infty$
achieves this goal, which 
 can be inferred with the help of 
 (\ref{intOmega0}) and (\ref{intOmega}).

Such a condition also guarantees
the UV convergence of the 
integral determining 
the norm of 
the wave-function.
This can be verified by combining 
 (\ref{normGeneral}) with (\ref{cohkap00}), 
 which leads to 
\be
1= \alpha^2 + \frac{\mu^2}{6\alpha^2\pi^2}
\int_0^\infty  d\om{k} \vareps{k}f^2(\om{k}),
\ \mu=|\boldmu|.
\label{hujik}
\ee
However, the discussed decay rate of $f$ 
does not guarantee 
the UV convergence of the 
expression for the energy
of the studied field configuration
because  
\be
{\cal H}=
\frac{\mu^2}{6\alpha^2\pi^2}
\int_0^\infty d\om{k}\vareps{k}^2f^2(\om{k})
\label{bjghvjh}
\ee
is obtained after  putting 
(\ref{HProca}) and (\ref{cohkap00}) into  (\ref{Hmean}).
Indeed, the requirement  of ${\cal H}<\infty$ 
implies that  
$f(\om{k})$ should vanish faster than 
$1/\om{k}^{3/2}$ for $\om{k}\to\infty$.

We introduce 
\be
 \gamma=2,4,6,\cdots
\label{g246}
\ee
and   settle for
\be
\label{fgamma}
f(\om{k})=\B{\frac{m}{\vareps{k}}}^\gamma,
\ee
which  fulfills the above requirement, 
obeys $f(\om{k}\to0)\to1$, is bounded, and allows 
us to re-use some 
technical results presented 
in \cite{BDPeriodic1}.

The integrals in (\ref{hujik}) and
(\ref{bjghvjh}) can 
be evaluated via
\be
\int_0^\infty 
d\om{k}\frac{\om{k}^{2a-1}}{\vareps{k}^{2b}}=
\frac{\Gamma(a)\Gamma(b-a)}{2m^{2(b-a)}\Gamma(b)}
\for b>a>0,
\label{calka}
\ee
which follows  from 
expression 
3.518.3 listed in \cite{Ryzhik}.
This results in   
\begin{subequations}
\begin{align}
&\alpha^2= \frac{1}{2}\B{1 \pm 
\sqrt{1-\B{\frac{\mu}{\mu_\text{max}}}^2}},\\
& \mu_\text{max}=\sqrt{\frac{3\pi^{3/2}\Gamma(\gamma-1/2)}{m^2\Gamma(\gamma-1)}},
\end{align}
\end{subequations}
\be
{\cal H}= 
m\frac{(m\mu)^2\Gamma(\gamma-3/2) }{12\alpha^2\pi^{3/2}\Gamma(\gamma-1)}.
\ee
Note that  $\mu_\text{max}$ 
provides the upper bound on
the 
magnitude of the 
magnetic  dipole moment that  can be encoded in the states 
discussed  in this section.

Having said all that, we are ready to
explore the non-equilibrium dynamics of the 
mean electromagnetic field. The 
computation of
the expectation values of (\ref{piE}) and 
(\ref{piB}) leads to  
\begin{align}
\label{MagcoE}
&\lara{\E(t,\rr)}=\Phi(t,r)\boldmu\times\rhat,\\
&\lara{\Bold(t,\rr)}=\Phi_+(t,r)(\boldmu\cdot\rhat)\rhat
- \Phi_-(t,r)  \boldmu,
\label{MagcoH}
\end{align}
where 
\begin{align}
\label{PhiII}
&\Phi(t,r)= \frac{1}{2\pi^2}
\int_0^\infty d\om{k} \om{k}\vareps{k} f(\om{k}) 
{\rm j}_1(\om{k}r) \sin(\vareps{k}t),\\
\label{PhipII}
&\Phi_+(t,r)=\frac{1}{2\pi^2}
\int_0^\infty d\om{k} \om{k}^2 f(\om{k}) 
{\rm j}_2(\om{k}r) 
\cos(\vareps{k}t),\\
&\Phi_-(t,r)=  \frac{1}{2\pi^2}
\int_0^\infty d\om{k} \om{k}^2 f(\om{k}) 
\BB{\frac{{\rm j}_1(\om{k}r)}{\om{k}r} - {\rm j}_0(\om{k}r)}
\cos(\vareps{k}t).
\label{PhimII}
\end{align}
These results have been obtained with the help of 
(\ref{intOmega0}), (\ref{intOmega}), and 
\be
\int d\Omega(\khat)k^i \exp(\ii \kk\cdot\rr)
\\= 4\pi  \ii \frac{\om{k}}{r}
{\rm j}_1(\om{k}r)r^i.
\ee

Expressions  (\ref{PhiII})--(\ref{PhimII})
suggest that the studied 
mean electromagnetic field is determined 
by  three
fairly non-trivial 
integrals  that 
cannot be found in references  
such  as \cite{Ryzhik}. 
However, we have been 
able to
show via standard manipulations that 
\begin{align}
\label{PhiRRR}
& \Phi(t,r)=\partial_t\partial_r\phi_\gamma(t,r),\\
& \Phi_{\pm}(t,r)=\partial_r^2\phi_\gamma(t,r)
\mp\frac{\partial_r\phi_\gamma(t,r)}{r},
\label{Phipm}
\end{align}
which suggests  that all we need to know here 
is the integral from (\ref{phiphi}).
This is only true as long as one can freely exchange 
the order of differentiation and integration
in the studied expressions. Namely, 
the equivalence of (\ref{PhiII})--(\ref{PhimII})
to (\ref{PhiRRR}) and  (\ref{Phipm}) 
relies upon the possibility of taking the 
derivatives, which 
are seen  in (\ref{PhiRRR}) and  (\ref{Phipm}),
under the integral sign in (\ref{phiphi}).
While there are no problems with doing so 
for $r,t>0$ and $\gamma=4,6,8,\cdots$, 
such a possibility should not be taken for granted
at $r=t$ when $\gamma=2$,
which is commented  upon in  Sec. \ref{subA}.

We will discuss now $\lara{\E(t,\rr)}$ and  
$\lara{\Bold(t,\rr)}$, 
mainly focusing our attention
on the mean magnetic field for the reason 
that will be soon evident.

\begin{figure}[t]
\includegraphics[width=\pref\columnwidth,clip=true]{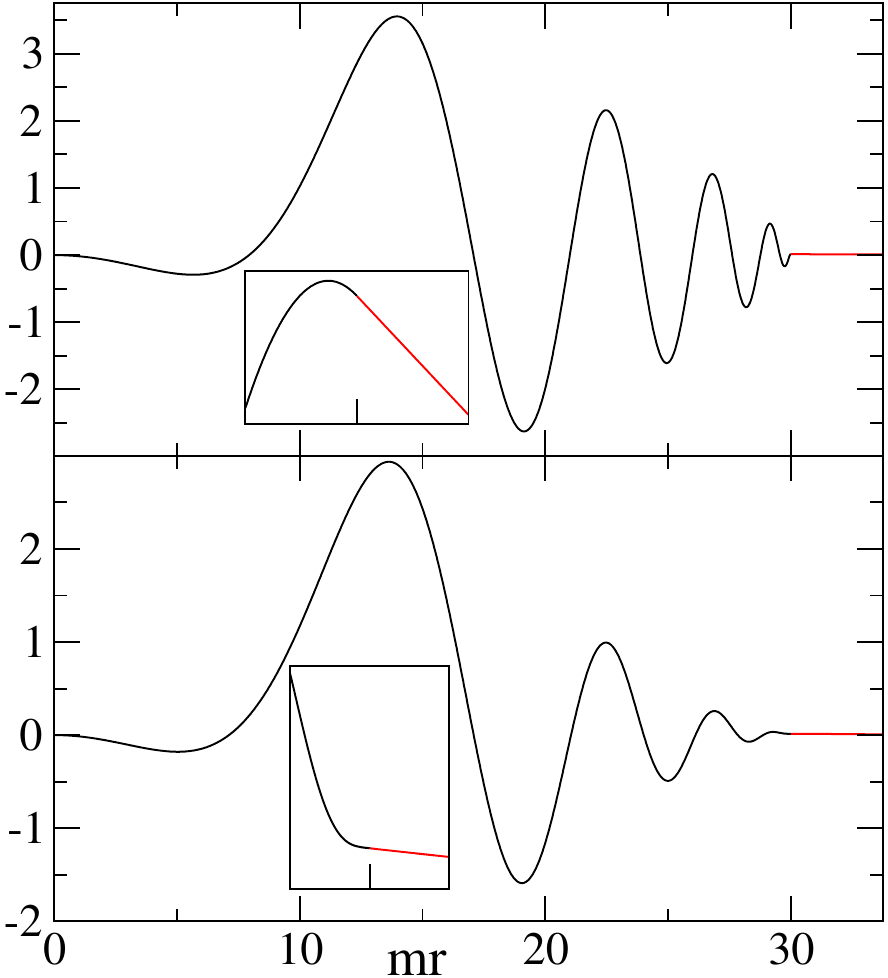}
\caption{
The rescaled coefficient $\Phi_+(t,r)$ for  $\gamma=4,6$.
Namely, $\Phi_+(t,r)\times10^4 m^{-3}$ at 
$m t=30$ for $\gamma=4$ (upper plot)
and $\gamma=6$ (lower plot).
Black (red)  lines show data for 
$r<t$ ($r>t$).
Insets  magnify the area around $r=t$.
Their  width in the
horizontal direction
is   $4\times10^{-4}$ (upper plot)
and $1$ (lower plot).
}
 \label{Phiplus46_fig}
\end{figure}

\begin{figure}[t]
\includegraphics[width=\pref\columnwidth,clip=true]{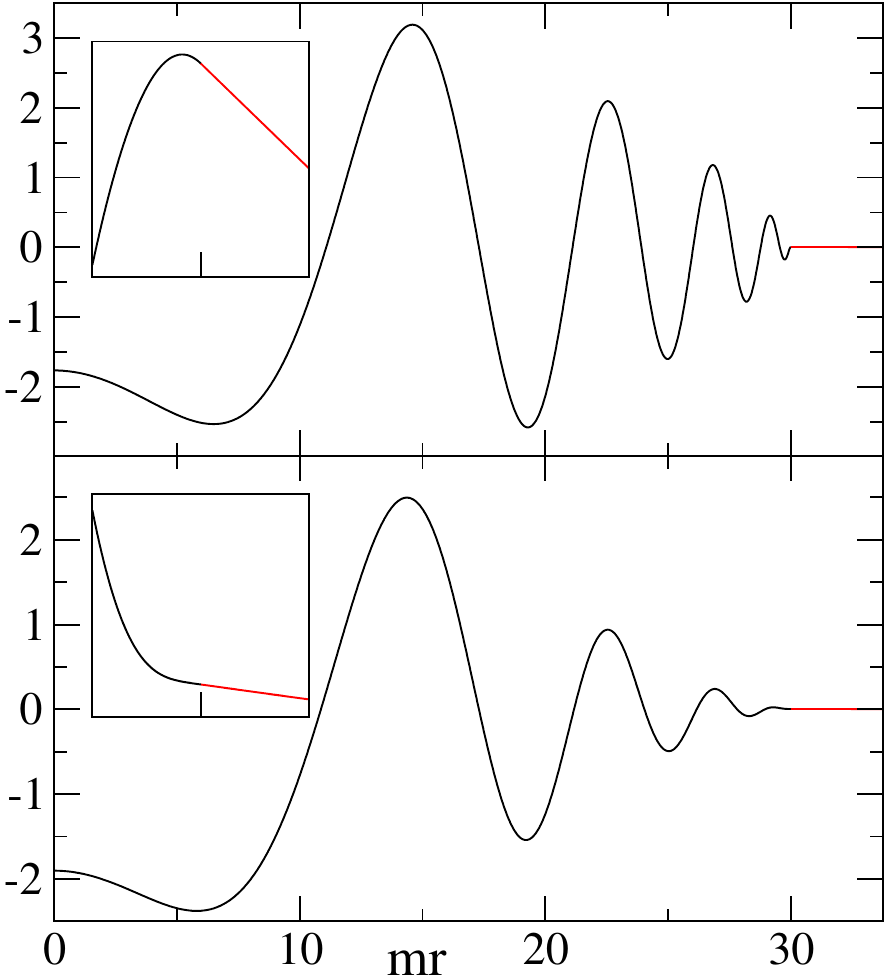}
\caption{The same as in Fig. \ref{Phiplus46_fig} except
we show  here $\Phi_-(t,r)$ instead of $\Phi_+(t,r)$.
The horizontal width of the insets 
is  $2\times10^{-4}$ (upper plot)
and $4\times10^{-1}$ (lower plot).
}
\label{Phiminus46_fig}
\end{figure}

First, 
the mean magnetic field 
for  $\gamma=4,6,8\cdots$ and $r\ge t$,
as well as $\gamma=2$ and  $r>t$, 
can be 
analytically obtained 
by putting  (\ref{hatphiLexp}) into 
(\ref{Phipm}) 
\begin{subequations}
\begin{align}
\label{BassM}
&\lara{\Bold(t,\rr)}= \frac{3(\boldmu\cdot\rhat)\rhat-\boldmu}{4\pi r^3}\cos( m t)
+ \exp(-mr) \boldsymbol{b}(t,\rr),\\
&\boldsymbol{b}(t,\rr) = 
\frac{\boldmu}{4\pi r^3}
\B{1+r\Delta_r+r^2\Delta_r^2}P_\gamma(mr,mt)
-\frac{3(\boldmu\cdot\rhat)\rhat}{4\pi r^3}
\B{1+r\Delta_r+\frac{r^2}{3}\Delta_r^2}P_\gamma(mr,mt),
\end{align}
\label{BrtMag}%
\end{subequations}
where $\Delta_r=m-\partial_r$.
These expressions show that 
   the  dynamics 
of the mean magnetic 
field,
in the $r>t$ region, is universal for $m t\gg1$.
Indeed,  
in such a case $\lara{\Bold(t,\rr)}$ practically 
does not depend on the parameter  $\gamma$,
which      is not 
uniquely specified in our studies (\ref{g246}).
The most important thing now is 
that  for $r\to\infty$,
we are left  with the first 
term in (\ref{BassM}), 
which  represents the 
magnetic field of the  
magnetic dipole having the
periodically oscillating  magnetic moment 
$\boldmu\cos( m t)$.

Second, 
the dynamics of  mean  magnetic 
field (\ref{MagcoH}) in the 
$r<t$ region 
is illustrated  
in Figs. 
\ref{Phiplus46_fig} and
\ref{Phiminus46_fig}, where we plot the 
 coefficients  $\Phi_\pm(t,r)$
for $\gamma=4,6$.
In order to prepare these figures, 
we have numerically 
evaluated   (\ref{PhipII}) and 
(\ref{PhimII}) via \cite{Mathematica131},
because we could  not arrive at  useful 
analytical 
expressions (in such a region  of space)  
for  $\gamma$'s greater than $2$ 
(the case of $\gamma=2$ is discussed 
in  Secs. \ref{subA}--\ref{subC}).
We see  from Figs. \ref{Phiplus46_fig} and
\ref{Phiminus46_fig} that the mean  magnetic field 
oscillates in  the $r<t$ region, where it
reverses 
its direction.
As we have numerically verified, 
the number (amplitude) of such oscillations 
increases (decreases) as a function of time. 
Note that such oscillations have a
 non-universal character because they 
depend on  $\gamma$.

Third,   the insets 
in Figs. \ref{Phiplus46_fig} and
\ref{Phiminus46_fig} illustrate 
the continuity of the mean magnetic 
field  across $r=t$ 
for $\gamma=4,6$ (such 
an observation also  holds for 
larger $\gamma$'s). 
However, the mean magnetic field is
weakly discontinuous at $r=t$ 
because there is a  shock wave 
propagating in the studied system.
Indeed,  the shock wave component  
of  $\Phi_\pm(t,r)$ becomes   evident  after 
the computation of
$\partial^2_t\Phi_\pm(t,r)$ for $\gamma=4$ 
and $\partial^4_t\Phi_\pm(t,r)$ for $\gamma=6$
(it can be shown with the help of 
\cite{BDPeriodic1}
that such derivatives are discontinuous 
at  $r=t$).
We mention in passing that
the numerical differentiation of the data 
from
Figs. \ref{Phiplus46_fig} and
\ref{Phiminus46_fig} supports the 
view that 
$\partial^2_r\Phi_\pm(t,r)$ for $\gamma=4$
and $\partial^4_r\Phi_\pm(t,r)$ for $\gamma=6$
are also discontinuous 
at  $r=t$.

\begin{figure}[t]
\includegraphics[width=\pref\columnwidth,clip=true]{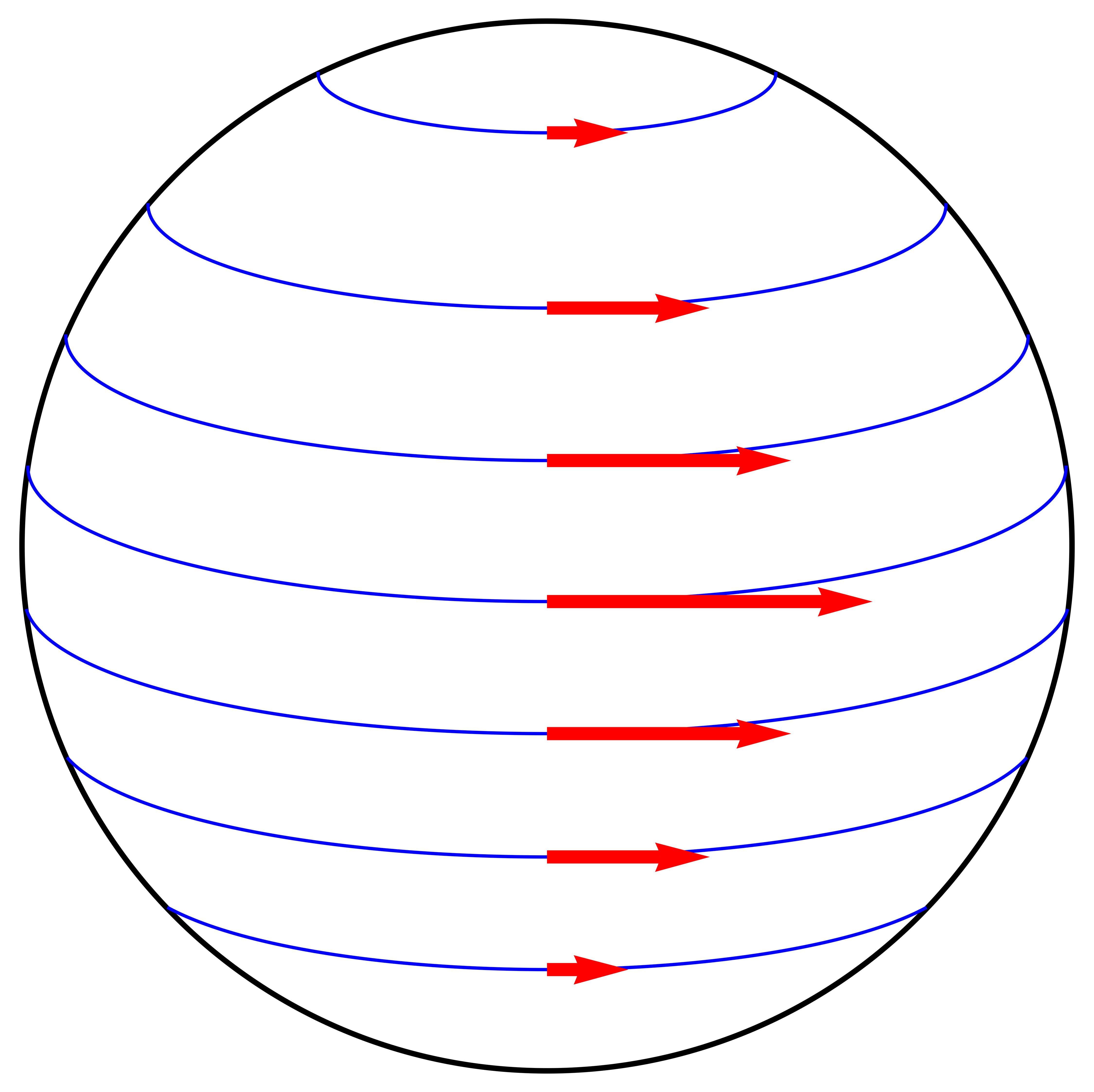}
\caption{Schematic plot of  mean electric field (\ref{MagcoE})
on the sphere of radius $r$.
 The  magnetic moment $\boldmu$ is 
oriented vertically (it points  upwards). 
The mean electric field   
$\lara{\E(t,\rr)}$ is presented with red arrows
at a given set of points ($\Phi(t,r)>0$ is assumed). 
It is symmetric with 
respect to  rotations around  the axis that 
is parallel to $\boldmu$ and goes through the center of 
the sphere. 
The field lines are depicted in blue, they are tangent 
to $\lara{\E(t,\rr)}$.
}
\label{rotation_fig}
\end{figure}

Fourth,
under the mapping 
$\boldmu\times\rhat \to q\rhat$,
$\lara{\E(t,\rr)}$ given by 
(\ref{MagcoE})
is equal to  
$\lara{-m^2\V(t,\rr)}$  computed in the
 states studied in  \cite{BDPeriodic1}. 
Such a   feature is rather unexpected
given the fact that the states discussed 
in this work  are 
of no interest in the context
of the problem considered in \cite{BDPeriodic1}.
Due to  the above-mentioned mapping, 
we shall not discuss below the dynamics of 
the mean electric field   in the magnetic
dipole-charged states. We only mention that 
for $r>t$ and $\gamma$'s given by (\ref{g246})
\be
\lara{\E(t,\rr)}= m \frac{\boldmu\times\rhat}{4\pi r^2}\sin( m t)
+ O\B{\exp(-mr)}.
\label{EmagLarge}
\ee
Such a formula predicts 
$|\lara{\E(t,\rr)}|\sim \sin(\theta)r^{-2}$ for $r\to\infty$,
where $\theta\in[0,\pi]$ is the angle between $\boldmu$
and  $\rhat$. We find it  interesting that despite the $r^{-2}$ 
asymptotic decay of 
$|\lara{\E(t,\rr)}|$, 
\be
\lara{ Q(t)}=\lim_{r\to\infty}\int d\boldsymbol{S}(\rr)\cdot\lara{\E(t,\rr)}=0
\label{Qzero}
\ee
in the discussed  states 
($d\boldsymbol{S}(\rr)$ is the surface element
on the sphere of radius  $r$).
Physically, (\ref{Qzero}) follows from the fact 
that in the Proca theory  one cannot 
construct  the state in which  $\lara{Q(t)}\neq0$ 
without  the  longitudinal excitations 
(such excitations are absent in the studied 
magnetic dipole-charged states).
Technically, (\ref{Qzero})  can be explained by the fact that 
 $\lara{\E(t,\rr)}$ is perpendicular to 
$d\boldsymbol{S}(\rr)$, which is so  not only
asymptotically (\ref{EmagLarge}) but also  for  any $r>0$
(\ref{MagcoE}).
Finally, we note that the  
axisymmetric topology of mean electric field (\ref{MagcoE}),
which is depicted in Fig. \ref{rotation_fig},
strikingly differs from the 
 topology of the 
Coulomb field despite the fact that 
such a   field  is  also falling off   as $r^{-2}$.
This fundamental  dissimilarity can be seen as the consequence 
of the difference in  the divergence of the two 
electric fields.
Namely, the divergence of
the Coulomb field is non-zero at its ``center'',
whereas $\lara{\DIV\E(t,\rr)}=0$ everywhere 
in the studied 
magnetic dipole charged-states because 
the divergence operator removes the transverse polarization modes from  
the electric field operator.

Fifth, we note that the mean electric and 
magnetic fields are perpendicular to 
each other, which is seen from 
(\ref{MagcoE}) and
(\ref{MagcoH}).\footnote{One may see this 
observation 
as the consequence of the fact that the mean 
vector field in Sec. \ref{Magnetic_sec}  is given by 
$\lara{V(t,\rr)}=\B{0,\lara{\V(t,\rr)}}$, where 
$\lara{\V(t,\rr)}\sim\boldmu\times\rhat$.
For such $\lara{V(t,\rr)}$, 
$-\partial_t\lara{\V(t,\rr)}$ and 
$\bnabla\times\lara{\V(t,\rr)}$
are perpendicular to each other
regardless of the form of the 
  function
providing the proportionality factor 
between $\lara{\V(t,\rr)}$ and $\boldmu\times\rhat$
(we assume differentiability of 
$\lara{\V(t,\rr)}$, which holds  
for $\gamma=4,6,8,\cdots$  as well 
as for $\gamma=2$ outside the shock wave front).
The above-stated $\lara{V^0(t,\rr)}=0$
can be seen as the consequence of the lack of
longitudinal excitations in
the magnetic dipole-charged
states, whereas  the proportionality 
of $\lara{\V(t,\rr)}$  to $\boldmu\times\rhat$
follows from the fact that 
$\lara{\V(t,\rr)}=-\partial_r\phi_\gamma(t,r)
\boldmu\times\rhat$ in such states. 
}
We also note  that the asymptotic 
algebraic decay of the mean electric field is 
slower than the one of the mean magnetic field,
which is seen from  (\ref{BrtMag}) and (\ref{EmagLarge}).

Till the end of this section,
we will discuss the 
$\gamma=2$ case, where one can obtain additional 
analytical 
insights into the non-equilibrium dynamics 
of the mean electromagnetic  field in  magnetic 
dipole-charged states.

\subsection{ $\gamma=2$: shock wave discontinuity}
\label{subA}
\begin{figure}[t]
\includegraphics[width=\pref\columnwidth,clip=true]{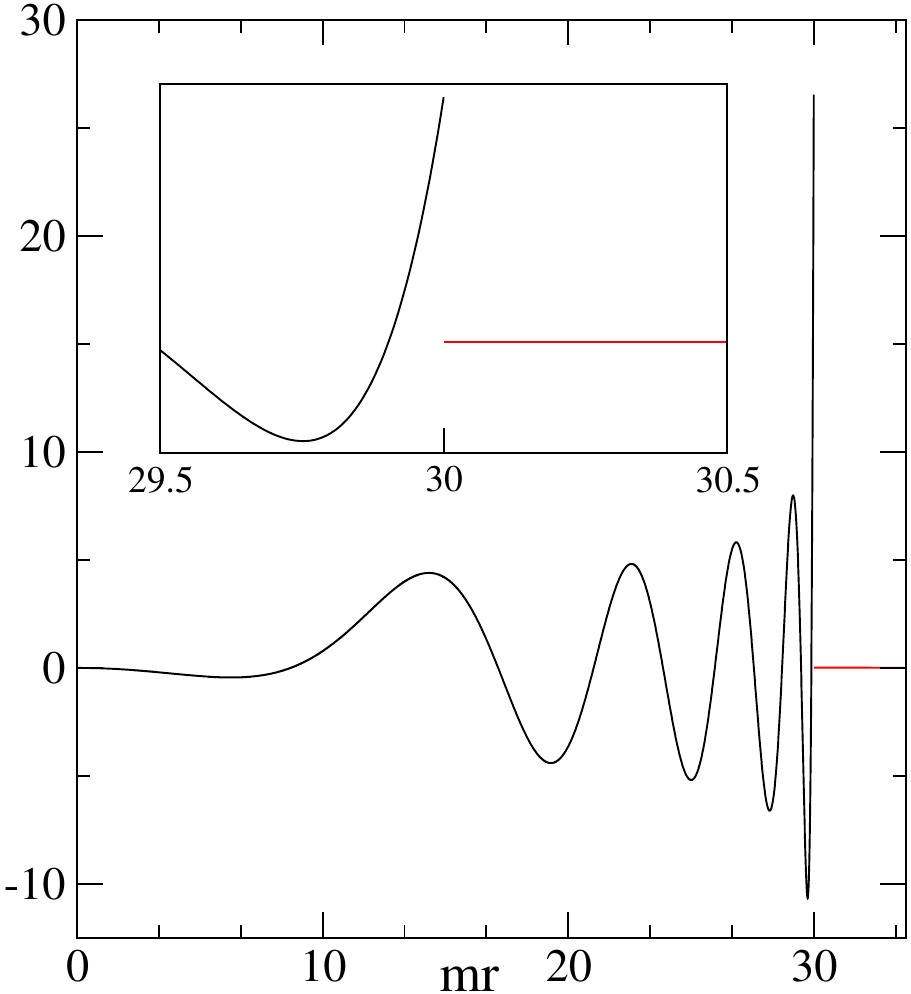}
\caption{The rescaled coefficient $\Phi_+(t,r)$ for  $\gamma=2$.
Namely, $\Phi_+(t,r)\times10^4 m^{-3}$ at 
$m t=30$.
The  black line comes from  (\ref{Bpm2}),
whereas the red one  from  (\ref{BpmAssym}).
The inset magnifies the area around the shock
wave discontinuity.
}
 \label{Bplus_fig}
\end{figure}

\begin{figure}[t]
\includegraphics[width=\pref\columnwidth,clip=true]{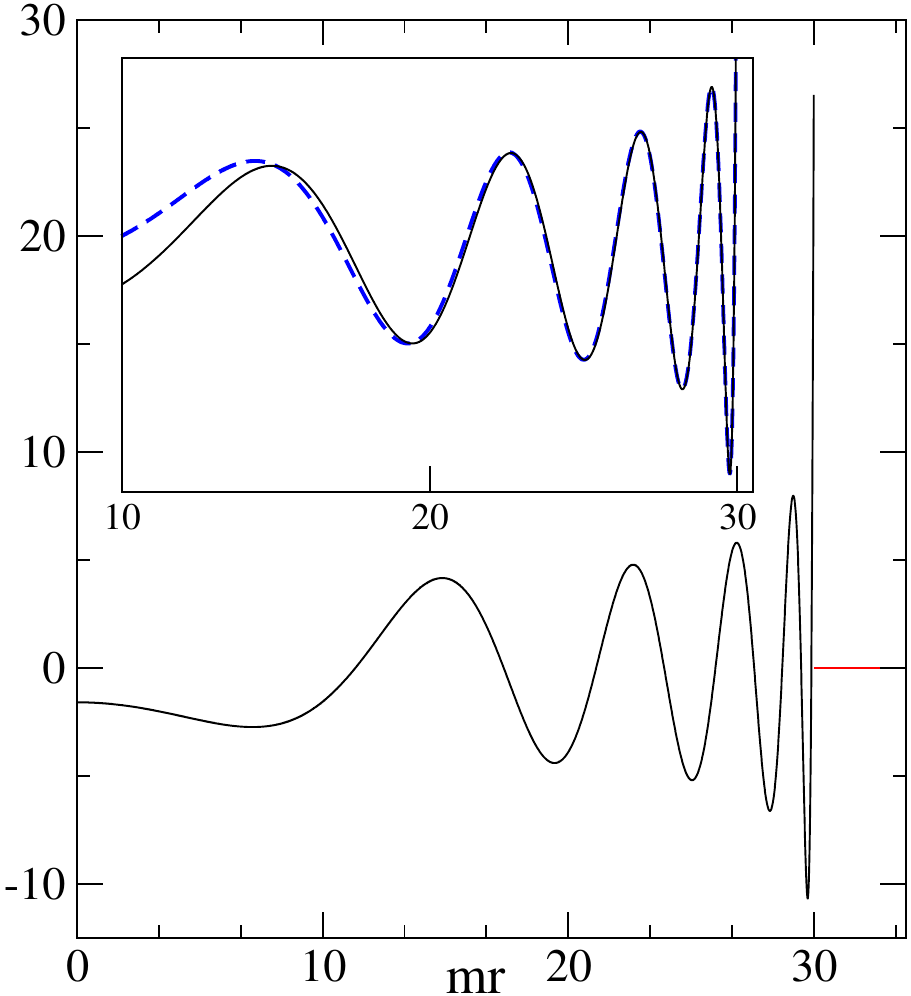}
\caption{Main plot: the same as in Fig. \ref{Bplus_fig} except
we  show here $\Phi_-(t,r)$ instead
of $\Phi_+(t,r)$.
Inset: comparision between the curves from the main plots 
of  Figs. \ref{Bplus_fig} (dashed blue line) 
and  \ref{Bminus_fig}  (solid black line)
in the $r<t$ region.
}
 \label{Bminus_fig}
\end{figure}

The key difference between the $\gamma=2$ case and 
$\gamma=4,6,8,\cdots$ cases is that
the mean magnetic field 
is discontinuous (continuous)
at  $r=t$  
 in the former  (latter) case(s).
Such a  discontinuity, which is illustrated in the insets of 
Figs. \ref{Bplus_fig} and
\ref{Bminus_fig},  can be explained as follows.

We   gather from  the results
presented in \cite{BDPeriodic1} that 
\be
\partial_r\phi_2(t,r)
= \partial_r\int_0^\infty d\om{k}\cdots
=\int_0^\infty d\om{k}\partial_r(\cdots)
\label{1stDER}
\ee
for any $r,t>0$, whereas 
\be
\partial^2_r\phi_2(t,r)
= \partial^2_r\int_0^\infty d\om{k}\cdots
 =\int_0^\infty d\om{k}\partial^2_r(\cdots)
\ee
for any $r,t>0$ as long as $r\neq t$.
These expressions imply that  for $r\neq t$,
the mean 
magnetic field can be computed from 
 (\ref{MagcoH}) combined with   (\ref{Phipm});
see the comments below (\ref{Phipm}). 
Thereby, to get insights into  $\lara{\Bold(t,\rr)}$
near $r=t$,  we  take a close look at 
$\partial_r\phi_2$ and $\partial^2_r\phi_2$.
Following  \cite{BDPeriodic1}, we
observe that 
\be
\lim_{r\to t^-} \partial_r\phi_2(t,r)
= \lim_{r\to t^+} \partial_r\phi_2(t,r),
\ee
whereas 
\be
\lim_{r\to t^-} \partial^2_r\phi_2(t,r)
= \frac{m^2}{4\pi t}
+ \lim_{r\to t^+} \partial^2_r\phi_2(t,r).
\label{d2ph2}
\ee
This implies that $\Phi_\pm(t,r)$ 
for $\gamma=2$ 
are discontinuous at  $r=t$ and so is  mean magnetic 
field (\ref{MagcoH}) 
\be
\lim_{r\to t^+}\lara{\Bold(t,r\rhat)}-
\lim_{r\to t^-}\lara{\Bold(t,r\rhat)}=
-\frac{m^2}{4\pi t} (\boldmu\cdot\rhat)\rhat+
\frac{m^2}{4\pi t} \boldmu.
\ee
In other words, there is a shock wave
discontinuity propagating with the
speed of light
in the discussed quantity.

The question now is what  can be said 
about the value of the mean magnetic field 
at the shock wave front. 
It turns out that there is a curious
ambiguity concerning this issue. Namely, 
\begin{subequations}
\begin{align}
\label{BBB3a}
\lara{\Bold(t,t\rhat)}&=\left.\lara{\bnabla\times\V(t,\rr)}\right|_{r=t}\\
&\neq
\left.\bnabla\times\lara{\V(t,\rr)}\right|_{r=t},
\label{BBB3b}
\end{align}
\end{subequations}
which triggers the following remarks.

First,   (\ref{BBB3a}) is given 
by (\ref{MagcoH}) with  $\Phi_\pm(t,r=t)$
obtained   from (\ref{PhipII})
and  (\ref{PhimII}).
For $\gamma=2$, the integrals in (\ref{PhipII})
and  (\ref{PhimII}) can be
calculated  with the help of
the results presented in \cite{BDPeriodic1}.
Indeed, by expressing 
 the
spherical Bessel functions
in terms of trigonometric 
functions, we have found  that 
 (\ref{PhipII})
and  (\ref{PhimII}) are given by 
  linear combinations of three integrals
that were computed 
in  \cite{BDPeriodic1}.
By following  such a procedure,
we have found that  
$\Phi_\pm(t,r=t)$ is given by
\be
\beta_\pm  \frac{ \cos(mt)-(1+mt)\exp(-mt) }{4\pi t^3}  
+m^2\frac{1/2-\exp(-m t)}{4\pi t},
\label{bbb3}
\ee
where  $\beta_\pm=2\pm1$.
This result turns out to be 
half-way between the  discontinuities
on  both sides of the shock wave front.
Namely,
(\ref{bbb3}) is equal to 
\be
\frac{1}{2} 
\lim_{r\to t^-}\Phi_\pm(t,r)+
\frac{1}{2}
\lim_{r\to t^+}\Phi_\pm(t,r),
\ee
which is seen from (\ref{BpmAssym}) and
(\ref{asdfghjklkjhgfd}).

Second, by computing the curl of 
\be
\lara{\V(t,\rr)}=-\partial_r\phi_2(t,r)
\boldmu\times\rhat,
\label{Vme}
\ee
one may verify  that (\ref{BBB3b})
is given
by  (\ref{MagcoH})  with  $\Phi_\pm(t,r=t)$
obtained from   (\ref{Phipm}).  
Such an expression is undefined 
because
$\partial_r^2\phi_2(t,r=t)$ does  not
exist \cite{BDPeriodic1}.
We mention in passing that 
(\ref{Vme}) can be  established 
for all $r,t>0$  with the help of
 (\ref{Vvecoperator}), 
(\ref{Oexp}),  
(\ref{cohkap00}), 
and (\ref{1stDER}).

Third, we expect that
(\ref{BBB3a}) answers the question 
of what is  the mean magnetic field 
at the  shock wave front. 
However, we lack a definite 
argument explaining why 
such a quantity should not  be 
approached  via (\ref{BBB3b}). Thereby, 
we leave open the issue discussed here.

Finally, we mention   that 
 similar observations apply to 
 mean electric field 
(\ref{MagcoE})
for $\gamma=2$. Such a quantity is also 
discontinuous at $r=t$, which can be linked 
to the fact that 
$\partial_t\partial_r\phi_2(t,r)$ is 
discontinuous there \cite{BDPeriodic1}.
Moreover, there is  an ambiguity in the 
evaluation of such a mean electric field
at  the shock wave front.
Namely,
$\lara{\E(t,t\rhat)}=\left.\lara{-\partial_t\V(t,\rr)}\right|_{r=t}$,
which is given by   (\ref{MagcoE}) combined with (\ref{PhiII}),
is not equal to  
$-\left.\partial_t\lara{\V(t,\rr)}\right|_{r=t}$
that  is given by 
(\ref{MagcoE}) combined with (\ref{PhiRRR}).
The former quantity is finite and it can be 
extracted out of \cite{BDPeriodic1} via the mapping 
stated above (\ref{Qzero}), whereas 
 the latter one is undefined because 
 $\partial_t\partial_r\phi_2(t,r=t)$
 does not   exist \cite{BDPeriodic1}.

\subsection{ $\gamma=2$: $r>t$ region }
\label{subB}

In the $r>t$ region, we find for  $\gamma=2$
that
\be
\label{BpmAssym}
\Phi_\pm(t,r)=\beta_\pm\frac{
\cos(mt)-(1+mr)\exp(-mr)}{4\pi r^3}
 - \frac{m^2}{4\pi r}\exp(-mr).
\ee
Such a result is 
obtained by 
 putting  (\ref{hatphiLexp}) into 
(\ref{Phipm}) and  noting  that
$P_2(a,b)=1$.
It determines the mean magnetic field,
in the considered region of space,
via (\ref{MagcoH}).

\subsection{ $\gamma=2$: $r<t$ region }
\label{subC}
To begin, we note that we 
know from \cite{BDPeriodic1} that 
for  $0<r<t$ 
\be
\phi_2(t,r)=\frac{\cos(mr)-\exp(-mr)}{4\pi r} 
-\frac{1}{4\pi r}
\int_{mr}^{mt} dy 
\int_0^{mr} dx
\rmJ_0\B{ \sqrt{y^2-x^2}},
\label{phi_small}
\ee
where 
$\rmJ_n$ is the Bessel function of the first kind
of order $n$.
This can be used to show that
for such $r,t$ and $\gamma=2$,
the coefficients from 
(\ref{MagcoH})  satisfy 
\begin{align}
4\pi r^3 \Phi_\pm(t,r)&=\beta_\pm 4\pi r\phi_2(t,r)-\beta_\pm m r \exp(-m r) 
+\beta_\pm m r \int_{mr}^{mt} dx \rmJ_0\B{\sqrt{x^2-(m r)^2}} \nonumber \\
&+(m r)^2[1-\exp(-m r)]
-(m r)^3\int_{mr}^{mt} dx \frac{\rmJ_1\B{\sqrt{x^2-(m r)^2}}}{\sqrt{x^2-(m r)^2}}.
\label{Bpm2}
\end{align}
We have obtained  this fairly complicated
result  via standard 
formulas associated with Bessel functions:
$\smallint_0^x dy{\rmJ_0}\B{\sqrt{x^2-y^2}}=\sin(x)$ coming 
from   formula 6.517 of \cite{Ryzhik}, 
$d{\rmJ_0}(x)/dx=-{\rmJ_1}(x)$, 
etc. The following remarks are in order now.

Near $r=0$, (\ref{Bpm2}) can be reduced to 
\begin{subequations}
\begin{align}
&\Phi_\pm(t,r) = \frac{m^3}{12\pi} (1\mp1)F(mt)+  O(r^2),\\
&F(x)= 1 + \rmJ_1(x) -x \rmJ_0(x)
-\frac{\pi x}{2}\rmJ_1(x)\rmH_0(x)
+\frac{\pi x}{2}\rmJ_0(x)\rmH_1(x),
\end{align}
\label{PhiPMPM}%
\end{subequations}
where  $\rmH_n$  represents
the Struve function   of order 
$n$.\footnote{The function $F(x)$ 
has been recently introduced 
in \cite{BDPeriodic1}. 
Fig. 5 of \cite{BDPeriodic1} illustrates
$F(x)$ for not-too-large $x$, whereas 
equation (71) of 
\cite{BDPeriodic1} can be used to 
justify equation (\ref{gggttt}) of this work (see also 
the remark below equation (70)
of \cite{BDPeriodic1}, which is useful 
in the derivation of equation (\ref{PhiPMPM}) of this work).}
We see   from (\ref{PhiPMPM})
that 
$\Phi_+(t,r=0)$ vanishes. Somewhat more 
interestingly, 
(\ref{PhiPMPM}) can be used  to show that 
$\Phi_-(t,r=0)$ exhibits damped oscillations,
which are accurately described for $mt\gg1$ by the formula
\be
\frac{2}{3}\B{\frac{m}{2\pi t}}^{3/2}
\sin\B{mt+\frac{\pi}{4}}.
\label{gggttt}
\ee

Near $r=t^-$, (\ref{Bpm2}) leads to 
\be
\lim_{r\to t^-}\Phi_\pm(t,r)=
\beta_\pm\frac{
\cos(mt)-(1+mt)\exp(-mt)}{4\pi t^3}
 +m^2\frac{1-\exp(-mt)}{4\pi t}.
\label{asdfghjklkjhgfd}
\ee
This implies that for $m t\gg1$,
the leading order contribution to 
$\Phi_+(t,r=t^-)$ and 
$\Phi_-(t,r=t^-)$
is the same.
In fact, as is illustrated  in the inset of 
Fig. \ref{Bminus_fig}, for $mt\gg1$
the two curves start to overlap for 
$r$ substantially smaller than $t^-$.
The $1/t$ decay of  $\Phi_\pm(t,r=t^-)$ 
   can be compared 
to the decay of $\Phi_\pm(t,r=t^+)$, which
according to (\ref{BpmAssym})
proceeds 
as $1/t^3$ (both results are valid
for $m t\gg1$).
 Thereby,  
the mean magnetic field
near   $r=t$ 
is dominated for $m t\gg1$   by the contribution from 
the region of space, which has just been swept by the 
shock wave front
(see e.g. the inset of  Fig. \ref{Bplus_fig}).

The dynamics of $\Phi_\pm(t,r)$ for $\gamma=2$
and $r<t$  is depicted in  Figs. \ref{Bplus_fig} and 
\ref{Bminus_fig}. 
To prepare them, 
we have numerically evaluated the integrals
from (\ref{Bpm2}) via \cite{Mathematica131}.
Apart from  the shock wave discontinuity,
these figures  are qualitatively similar to Figs. 
\ref{Phiplus46_fig} and
\ref{Phiminus46_fig}, which we have  just  
discussed. Therefore, we shall not dwell on them.

\section{Electric dipole-charged states}
\label{Electric_sec}

The states of interest here will be
constructed so as to 
yield $\lara{\Bold(0,\rr)}=\0$ for $r\ge0$ and 
\be
\lara{\E(0,\rr)}=
\frac{3(\ddd\cdot\rhat)\rhat-\ddd}{4\pi r^3}
\label{E00d}
\ee
for $r\to\infty$, where
$\ddd$ is the electric dipole moment.

To achieve this goal, we consider  the ansatz 
\be
\chi(\xx) =\int \frac{\dd{k}}{(2\pi)^{3/2}}
g(\kk) a_{\kk3} \exp(\ii\kk\cdot\xx)+\hc,
\ee
where $g(\kk)\in\RR$, 
which via (\ref{piE}), (\ref{piB}), and  (\ref{Oexp}) leads to $\lara{\Bold(0,\rr)}=\0$
and 
\be
\lara{\E(0,\rr)}=\frac{1}{2}
\int \frac{\dd{k}}{(2\pi)^3}\sqrt{\frac{2}{\vareps{k}}}
\frac{m}{\om{k}}g(\kk)\kk\exp(\ii\kk\cdot\rr) + \cc
\label{E0000}
\ee
If we substitute 
\be
g(\kk)=-\sqrt{\frac{\vareps{k}}{2m^2}}\frac{\ddd\cdot\kk}{\om{k}}
\ee
into (\ref{E0000}), replace $(\ddd\cdot\kk)\kk$ by 
$-(\ddd\cdot\bnabla)\bnabla$, and
uncritically reverse the order of 
differentiation and integration, 
we  find  that the resulting expression
is the same as (\ref{E00d}) for all $r>0$. 
Such a procedure of the evaluation of 
$\lara{\E(0,\rr)}$, however, 
is unjustified. In fact, 
as can be easily verified
with the help of (\ref{intOmega}),
the situation here is  
analogous to the one 
 discussed in Sec. \ref{Magnetic_sec}. 
Thereby, it should come as no 
surprise that we again 
introduce the bounded function $f(\om{k})\in\RR$ 
and proceed via
\be
\chi(\xx)\to\chi(\xx) =
-\int \frac{\dd{k}}{(2\pi)^{3/2}}\sqrt{\frac{\vareps{k}}{2 m^2}}
\frac{\ddd\cdot\kk}{\om{k}}f(\om{k})
a_{\kk3} \exp(\ii\kk\cdot\xx)+\hc,
\ee
where $f(\om{k})$ vanishing faster 
than $1/\om{k}$ for $\om{k}\to\infty$
ensures  the UV convergence of the integral 
determining  $\lara{\E(0,\rr)}$, 
whereas $f(\om{k}\to0)\to1$ protects 
the proper asymptotic form of 
such a mean electric field.

\begin{figure}[t]
\includegraphics[width=\pref\columnwidth,clip=true]{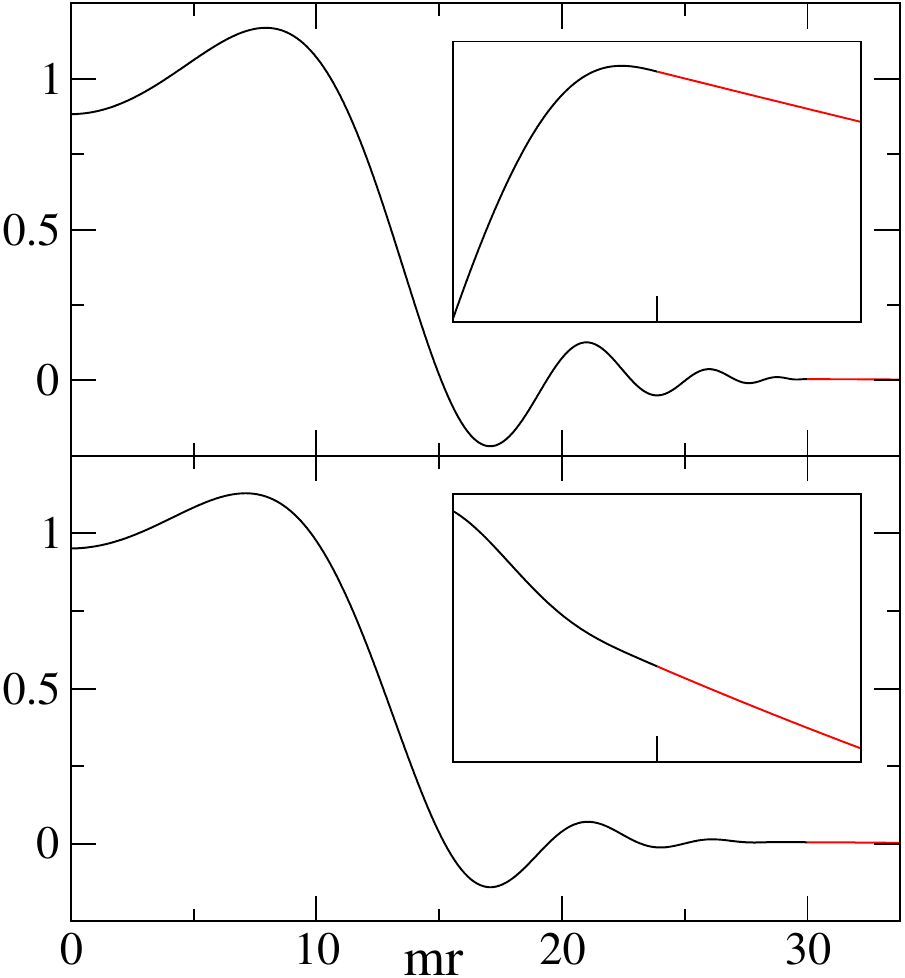}
\caption{
The rescaled coefficient $\frac{1}{2}[\Phi_+(t,r) -  \Phi_-(t,r)]$ for $\gamma=4,6$.
Namely, 
$\frac{1}{2}[\Phi_+(t,r) -  \Phi_-(t,r)]\times10^4 m^{-3}$ at 
$m t=30$ for $\gamma=4$ (upper plot)
and $\gamma=6$ (lower plot).
Black (red)  lines show data for 
$r<t$ ($r>t$).
Insets  magnify the area around $r=t$.
Their  width in the
horizontal direction
is   $4\times10^{-1}$ (upper plot)
and $2$ (lower plot).
}
\label{electric_fig}
\end{figure}

Further constraints on $f$ come from   
(\ref{normGeneral}) and (\ref{Hmean})
leading to 
\be
1= \alpha^2 + \frac{d^2}{12\alpha^2\pi^2 m^2}
\int_0^\infty  d\om{k}\om{k}^2\vareps{k}f^2(\om{k})
\label{eq111}, \ d=|\ddd|
\ee
and
\be
{\cal H}=
\frac{d^2}{12\alpha^2\pi^2 m^2}
\int_0^\infty d\om{k}\om{k}^2\vareps{k}^2f^2(\om{k}),
\label{eq222}
\ee
which substantially     differ from 
(\ref{hujik}) and (\ref{bjghvjh}),
respectively.
In fact, 
$f(\om{k})$ vanishing faster than $1/\om{k}^{5/2}$
for  $\om{k}\to\infty$  has to be assumed now. 
We again choose  
$f(\om{k})$ given by (\ref{fgamma}).
This time, however, we consider 
\be
\gamma=4,6,8,\cdots.
\label{g468}
\ee
The evaluation of (\ref{eq111}) and 
(\ref{eq222}) via (\ref{calka})
leads to 
\begin{subequations}
\begin{align}
&\alpha^2= \frac{1}{2}
\B{1 \pm 
\sqrt{1-\B{\frac{d}{d_\text{max}}}^2}},\\
&d_\text{max}= 
\sqrt{\frac{12\pi^{3/2}\Gamma(\gamma-1/2)}{m^2\Gamma(\gamma-2)}},
\end{align}
\end{subequations}

\be
{\cal H}=
m\frac{(md)^2\Gamma(\gamma-5/2)}{48\alpha^2\pi^{3/2}\Gamma(\gamma-1)},
\ee
where $d_\text{max}$ provides the upper bound on the 
magnitude of the 
electric dipole moment that can be encoded in the states 
studied  in this section.

The  field configuration, in the discussed 
electric-dipole charged states, is 
characterized by 
\begin{align}
\label{ElcoE}
&\lara{\E(t,\rr)}=\Phi_+(t,r)(\ddd\cdot\rhat)\rhat
-\frac{\Phi_+(t,r)-\Phi_-(t,r)}{2}\ddd,\\
&\lara{\Bold(t,\rr)}=\0,
\label{errdrtyu}
\end{align}
where $\Phi_\pm(t,r)$ are given by (\ref{PhipII}) 
and (\ref{PhimII}).
Several remarks are in order now.

First,  we note 
that for  $\gamma$'s
 being of interest here (\ref{g468}), 
one  may  equivalently 
evaluate   (\ref{ElcoE}) via 
 (\ref{Phipm})  for all $\protect{r,t>0}$.
 We also note that 
 mean electric field (\ref{ElcoE})
 is weakly  discontinuous at $r=t$
for such   $\gamma$'s.

Second, by combining (\ref{ElcoE}) with 
(\ref{Phipm}) and (\ref{hatphiLexp}), 
we have found that 
for $r\ge t$ and $\gamma$ given by (\ref{g468})
\begin{subequations}
\begin{align}
\label{EelLargeA}
&\lara{\E(t,\rr)}=
\frac{3(\ddd\cdot\rhat)\rhat-\ddd}{4\pi r^3}
\cos( m t) +\exp(-mr) \boldsymbol{e}(t,\rr),\\
&\boldsymbol{e}(t,\rr)=
\frac{\ddd}{4\pi r^3}
(1+r\Delta_r)P_\gamma(mr,mt)
-\frac{3(\ddd\cdot\rhat)\rhat}{4\pi r^3}
\B{1+r\Delta_r+\frac{r^2}{3}\Delta_r^2}P_\gamma(mr,mt).
\end{align}
\label{EelLarge}%
\end{subequations}
The  hallmark feature of 
such a solution is that 
in the limit of large $r$,
we are left in (\ref{EelLargeA})
with the field of the
 electric dipole
having the  periodically
oscillating  dipole moment
$\ddd\cos( m t)$.

Third, the dynamics of the  coefficient in the 
first term in 
(\ref{ElcoE}) is depicted in Fig. 
\ref{Phiplus46_fig} for   $\gamma$'s that are 
 of interest here.
The dynamics of the coefficient in the 
second term in 
(\ref{ElcoE}) is 
illustrated in Fig.
\ref{electric_fig},
which is directly  related to 
Figs. \ref{Phiplus46_fig} 
and \ref{Phiminus46_fig}.
Thereby, we shall not dwell on it.

Fourth,  mean electromagnetic field 
(\ref{ElcoE}) and 
(\ref{errdrtyu}) 
  is interesting
  from the Maxwell theory perspective.
 Namely,  due to the 
identity\footnote{Equations
(\ref{gtyty})      and (\ref{hffr}) are written under the 
      tacit assumption that  there are no problems
      with the  differentiation of $\lara{\E}$ and 
      $\lara{\Bold}$.}
\be
\bnabla\times\lara{\E} =- \partial_t\lara{\Bold},
\label{gtyty}
\ee
which holds not only in the Maxwell theory but
also in the Proca theory,
one may guess that (\ref{ElcoE}) can be written as
the  gradient of a scalar because the right-hand side
of (\ref{gtyty}) vanishes due to  (\ref{errdrtyu}). 
This is indeed the case as it turns out that 
 (\ref{ElcoE}) can be also expressed 
 as
\be
\lara{\E(t,\rr)}=\bnabla[\ddd\cdot\bnabla\phi_\gamma(t,r)].
\ee
Then, we note that the  
time-dependence of
the mean electric field,
in the presence of the 
 vanishing mean  magnetic field, 
 suggests 
 the existence of the  mean current 
 in our system. 
Such a suggestion, based on the 
intuition coming from   the  Maxwell 
theory, appears to be confusing 
at first sight because there is no
{\it external} current
in our calculations.
However, it 
turns out to be correct
 because  in 
 the Proca theory
 \be
 \bnabla\times\lara{\Bold}=\lara{\J} +\partial_t\lara{\E},
 \label{hffr}
 \ee
 where $\J=-m^2\V$ is the {\it internal} 
 $3$-current
 operator of theory (\ref{SPr})   
(see  \cite{BDPeriodic1} for its 
recent discussion in the context relevant for  
these studies). 
We mention in passing that the  classical internal 
Proca current, $-m^2\V_\text{cl}$,
is discussed   in \cite{Nieto_RMP1971,Nieto_RMP2010}.
Such a quantity, despite the fact that it is 
proportional to the vector field $\V_\text{cl}$, is 
observable   in the Proca theory
(e.g. see the discussion below equation (2.19) of  \cite{Nieto_RMP1971}
and reference  \cite{LakesPRL1998} reporting 
an interesting attempt of 
the experimental measurement of $\V_\text{cl}$).

\section{Summary}
\label{Summary_sec}

We have discussed the dynamics of field configurations 
encoded in the  particular  
class of electric and  magnetic  dipole-charged states
in the Proca theory.
The key universal features of our 
results 
can be transparently presented 
by taking a look at the asymptotic fields
\begin{align}
\label{EasE}
\lara{\E(t,\rr)}_\infty =\lara{\E(t,\rr)} \for r\to\infty, \\ 
\lara{\Bold(t,\rr)}_\infty =\lara{\Bold(t,\rr)} \for r\to\infty,
\label{BasB}
\end{align}
which are obtained by discarding  the 
short-distance 
  terms in  (\ref{BrtMag}), 
 (\ref{EmagLarge}), and (\ref{EelLarge}). 
In our magnetic  dipole-charged states
\begin{align}
\label{Msola}
& \lara{\Bold(t,\rr)}_\infty  = 
\frac{3(\boldmu\cdot\rhat)\rhat-\boldmu}{4\pi r^3}\cos( m t),\\
& \lara{\E(t,\rr)}_\infty =
m \frac{\boldmu\times\rhat}{4\pi r^2}\sin( m t),
\label{Msolb}
\end{align}
whereas in our   electric  dipole-charged states
\begin{align}
\label{Esol}
&\lara{\E(t,\rr)}_\infty =
\frac{3(\ddd\cdot\rhat)\rhat-\ddd}{4\pi r^3}\cos( m t), \\
&  \lara{\Bold(t,\rr)}_\infty  =\0.
\label{Bsol}
\end{align}
The first thing we learn from  these results  is that 
the asymptotic fields satisfy the harmonic oscillator
equation
\be
\partial^2_t \lara{O(t,\rr)}_\infty = -m^2 
\lara{O(t,\rr)}_\infty,
\ee
where $O=\E,\Bold$.
The second one is that  according 
to (\ref{Msola}) [(\ref{Esol})], 
the studied   magnetic [electric] dipole-charged 
field configurations are  
characterized by the periodically 
oscillating magnetic [electric] dipole moment
$\boldmu\cos(mt)$ [$\ddd\cos(mt)$]. 
Therefore, our  results provide the 
concrete
theoretical 
illustration of  the 
phenomenon of the  {\it 
periodic oscillations of the 
dipole moments} in the Proca theory, 
which to the best of our 
knowledge has not been discussed 
in the literature before.

We have also discussed 
the non-equilibrium dynamics of 
our electric and magnetic 
dipole-charged field configurations 
at intermediate distances, 
where the  shock wave phenomenon seems to be most interesting. 
Namely,  our solutions
   are either discontinuous or  
weakly discontinuous and these
discontinuities are propagating.
We find it interesting that such singularities  
appear despite the fact that the studied quantum 
states are well-defined. Indeed, 
they are normalizable and represent finite-energy 
field configurations.

The non-equilibrium character 
of our solutions stems from the fact that there is 
no external current keeping the fields  in place.
As a result of that, the discussed  fields escape from their 
initial arrangement. In other words, 
 one deals  with  escaping (outgoing)
solutions   when our results 
are considered in the time domain 
$t:0\to\infty$, which has been  conveniently 
chosen for the discussion  of
the dipole-charged states in 
Secs. \ref{Magnetic_sec} and \ref{Electric_sec}. 
However, one can 
also consider our solutions in any time
domain, say  $t:-\infty\to\infty$.
Then for  $t:-\infty\to0$, 
one finds the dynamics quite   similar to the one 
discussed  in Secs. \ref{Magnetic_sec} and \ref{Electric_sec}.
It is so because our solutions are
either symmetric or antisymmetric 
with respect to the $t\to-t$ transformation,
which is seen from 
(\ref{MagcoE}),
 (\ref{MagcoH}), (\ref{ElcoE}), and
 (\ref{errdrtyu}).
Thereby, in the time domain $t:-\infty\to0$,
one deals with collapsing (incoming) solutions. 
In particular, when $t:-\infty\to\infty$, 
 the above-mentioned singularities 
first collapse towards the point $\rr=\0$ and then
they escape  to spatial infinity.

The physical realization of the discussed states 
is problematic because 
(i) it is unclear what stable particle
could be  described by the Proca theory
and (ii)  causality 
considerations  prohibit the
laboratory-based creation of the dipole-charged 
states.
Regarding the (i) issue, we mention that 
it is still  possible that the photon 
is a massive 
particle \cite{Nieto_RMP1971,Gillies2005,Nieto_RMP2010};
other options may arise in the future. 
Regarding the (ii) issue, we mention that 
the perpetual evolution  scenario, where the evolution starts
at  $t=-\infty$ and the fields undergo
initially collapsing dynamics, 
somewhat  avoids the 
problem with the experimental creation of 
the dipole-charged states.

Finally, we would like to say that our studies
give  definite  insights into the structure and 
dynamics of the  IR sector of the Proca theory.
This is a fairly unexplored topic because 
 short range fields are traditionally
associated with such a theory.
In particular, the generalization 
of our results to all multipole 
field configurations can be considered 
in the future studies.
We would like to stress that
the characterization of the IR sector of the
Proca theory poses  a  
well-defined mathematical problem
and it  contributes  to the in-depth 
understanding of such a paradigmatic
theory of a  massive vector field.

\section*{ACKNOWLEDGMENTS}
These studies have
been  supported by the Polish National
Science Centre (NCN) Grant No. 2019/35/B/ST2/00034.
The research for this publication has been also supported 
by a grant from the Priority Research Area DigiWorld under
the Strategic Programme Excellence Initiative at Jagiellonian University.

\appendix
\section{CONVENTIONS}
\label{Conv_app}

We adopt  the Heaviside-Lorentz system of units
in its $\hbar=c=1$ version.
Greek and Latin indices of tensors  take values $0,1,2,3$ and   $1,2,3$,
respectively. 
The metric signature is $(+---)$.
$3$-vectors are written in bold, e.g. $x=(x^\mu)=(x^0,\xx)$.
We use the Einstein summation convention,  
$(X_{\mu\cdots})^2=X_{\mu\cdots}X^{\mu\cdots}$.
$\partial_X=\partial/\partial X$
and  $X^+$ ($X^-$) denotes
the quantity that is infinitesimally larger (smaller)
than $X$. The hermitian (complex)
conjugation is denoted as $\hc$ ($\cc$).

\section{$P_\gamma$ polynomials}
\label{Polynomial_app}

The following formula for the $P_\gamma$ polynomials,
which we have introduced in this work in 
 (\ref{hatphiLexp}),
was  given in  \cite{BDPeriodic1}
\begin{subequations}
\begin{align}
& P_\gamma(a,b) =  -\exp(a)
\text{Res}(f(z),\ii\pi/2),\\
& f(z)= \frac{\cos[b\cosh(z)]
\exp[\ii a\sinh(z)]}{\sinh(z)\cosh^{\gamma-1}(z)},
\end{align}
\label{IAB}%
\end{subequations}
where $\gamma=2,4,6,\cdots$ and 
${\rm Res}(f(z),z_0)$ stands for the residue of the function 
$f(z)$ at $z_0$. Given the fact that $f(z)$ has 
the pole of 
order $\gamma-1$ at $z=\ii\pi/2$, the above 
expression is fairly complicated.
We will derive   another  
formula below, the one  
allowing for  the  recursive evaluation of
$P_\gamma(a,b)$.

To begin, we introduce dimensionless variables 
\be
a=m r, \ b=mt
\ee
and the following  function
\be
\widehat{\phi}_\gamma(a,b)= 4\pi r \phi_\gamma(t,r) 
= \frac{2}{\pi} \int_0^\infty
d\omega \frac{ \cos\B{b\sqrt{1+\omega^2}} }{(1+\omega^2)^{\gamma/2}}   
\frac{\sin(a \omega )}{\omega},
\label{whatINT}
\ee
which for $a\ge b\ge 0$  reduces to 
\cite{BDPeriodic1}
\be
\widehat{\phi}_\gamma(a,b)=
\cos(b)-P_\gamma(a,b)\exp(-a).
\label{what1}
\ee
Note that unlike $\widehat{\phi}_\gamma(a,b)$,
the  $P_\gamma(a,b)$ polynomials are 
insensitive to the relation between $a$ and 
$b$, which is seen  from  (\ref{IAB}).

For  
\be
\gamma=4,6,8,\cdots
\ee
being   of 
interest from now on,
we find that 
\be
\partial_b^2\widehat{\phi}_\gamma(a,b)
=-\widehat{\phi}_{\gamma-2}(a,b).
\label{what3}
\ee
Such an equation follows 
from the fact that
according to \cite{BDPeriodic1},  
$\partial_b^2$ can be taken under the 
integral symbol in (\ref{whatINT})
during the evaluation of the 
left-hand side of (\ref{what3}).

By combining (\ref{what1}) with (\ref{what3}), 
we arrive at
\be
\label{deq1}
\partial_b^2P_\gamma(a,b)=-P_{\gamma-2}(a,b),
\ee
whose solution can be written as  
\be
P_\gamma(a,b)=P_\gamma(a,0) -
\int_0^b dy \int_0^y dx P_{\gamma-2}(a,x).
\label{Pga}
\ee
Note that the double integration of 
(\ref{deq1}) over $b$
does not lead to the term linear in  
$b$ because 
$P_\gamma(a,b)$ is  even in $b$, which follows
from   (\ref{IAB}).

The expression for 
 $P_\gamma(a,0)$
can be directly obtained from 
the $b=0$ version of 
(\ref{whatINT}). Namely, 
formula 3.737.3 of \cite{Ryzhik} 
yields 
\be
P_{\gamma}(a,0)=
P_{\gamma-2}(a,0)+
\frac{a}{\gamma-2}\B{1-\frac{d}{da}}P_{\gamma-2}(a,0).
\label{PgamRec}
\ee
By combining (\ref{Pga}) with (\ref{PgamRec}), 
(\ref{PPrec}) is established.


%

\end{document}